\documentclass[12pt]{article}
\usepackage{color}
\usepackage{graphicx}
\usepackage{amssymb}
\usepackage{breqn}
\usepackage[font=small,labelfont=bf]{caption}
\usepackage{float}

\title{\textcolor{black}{ \textbf{Semi-analytical derivation of the 2D all-FLR ICRH wave equation as a high-order partial differential equation}}}  
\author{D. Van Eester  \& E.A. Lerche\\ 
Laboratory for Plasma Physics \\
EUROfusion Consortium \& Trilateral Euregio Cluster Member\\
Renaissancelaan 30 Avenue de la Renaissance\\
B-1000, Brussels, Belgium \\
\textcolor{black}{\small{\textit{corresponding author: Dirk Van Eester (email: d.van.eester@fz-juelich.de)}}}\\
}
\begin{document}

\maketitle

\begin{abstract}
\textcolor{black}{For 1-dimensional applications,} Bud\'{e}'s method \cite{Bude} \textcolor{black}{has been shown to be capable of accurately solving} the all-FLR (Finite Larmor Radius) integro-differential wave equation as a high-order differential equation allowing to represent all \textit{physically relevant} (fast, slow and Bernstein) modes upon making a polynomial fit that is accurate in the relevant part of $\vec{k}$ space. \textcolor{black}{The adopted fit is superior to the Taylor series expansion traditionally adopted to truncate the series of finite Larmor radius corrections, while the differential rather than integro-differential approach allows for significant gain in required computational time when solving the wave equation.} 
The method was originally proposed and successfully tested in 1D for radio frequency (RF) waves and in absence of the poloidal field \cite{AllFLRBudeDVE}. In the present paper, the derivation of the extension of that procedure to 2D and for finite poloidal field - semi-analytically yielding the coefficients of the relevant high-order partial differential equation - is discussed in preparation of future numerical application. 
\end{abstract}

\section{Introductory note}

In the past decades, many authors have devoted time to finding speedy but accurate ways to solve the wave equation reigning the wave propagation and damping in magnetic confinement machines. A number of bottlenecks were encountered and - at least partially - solved. 
\begin{itemize}
\item The first is that finite temperature (and hence finite Larmor radius) effects are crucial to understand wave damping and to determine how the wave polarisation - itself a key quantity in determining the absorption efficiency - changes under the influence of the presence of the plasma. Traditionally, a Taylor series expansion in $k_\perp\rho_L$ (where $k_\perp$ is the perpendicular wave number and $\rho_L$ is the Larmor radius) truncated at second order terms is used \cite{Fukuyama}, \cite{LamalleThesis}, \cite{Brambilla}, \cite{DumontEVE}. Although perfectly suitable for describing the fast magnetosonic wave in the ICRH (Ion Cyclotron Resonance Heating) domain for plasmas at moderate temperature, it is insufficient to correctly describe the fate of short wavelength branches (in the RF domain typically the ion Bernstein wave excited at the confluence near the ion-ion hybrid layer and hence routinely present at or close to the main damping region). Moreover, since RF heating typically creates high-energy tails and since fusion-born populations typically are highly energetic, the small-FLR assumption's validity needs to be checked a posteriori even for the fast wave in many relevant conditions. Luckily, the typically exploited truncation allows to at least get a first idea of what the actual wave absorption's strength is for most experimentally exploited regimes. But a more general procedure is nevertheless highly desirable. To date, the AORSA code \cite{AORSA} is likely the most general \textcolor{black}{ICRH} wave solver available that does not suffer from the small-FLR assumption routinely made. Both its strength and weakness is that it solves the all-FLR integro-differential wave equation by writing the electric field in its Fourier form but replaces the continuous integrals over $\vec{k}$ space by discrete sums. This makes the code very general but requires massive \textcolor{black}{computer} time and memory to invert the full matrix resulting from retaining the couplings between all modes in the $\vec{k}$ spectrum. Cutting down on \textcolor{black}{computational} needs is a useful task while keeping the description as general as possible is a worthwhile exercise.  
\item In inhomogeneous plasmas, the wave vector is not a constant and hence dispersion equation solving is insufficient to get a full grip on the wave dynamics. Because of the poloidal as well as toroidal periodicity of magnetic confinement devices, Fourier expansions are often used to describe the dynamics in the periodic directions and an associated set of variables $(\rho,\theta,\varphi)$ - where $\rho$ is the minor radius, $\theta$ is the poloidal and $\varphi$ the toroidal angle -  is routinely adopted. This introduces an artificial singularity in the associated coordinate system: although the magnetic axis in a tokamak ($\rho=0$) physically is a perfectly regular point, the mathematical singularity introduced requires one to carefully treat the region close to the axis, in particular when kinetic effects are important so that the wave flux is not only carried electromagnetically but also by particles in coherent motion with the waves. For an example of the evaluation of the kinetic flux - be it in 1-D - see e.g. \cite{TOMCAT,TOMCAT-U1}. Physically nor the Poynting nor the kinetic flux has to be zero at the magnetic axis; the total flux vector should simply be continuous when crossing this point. 
For finite element representations where the weak formalism is exploited - and to the exception of finite element formulations relying on base functions for which the continuity across individual finite element borders is automatically guaranteed by construction (see the brief section on fluxes) - 
that requires the kinetic flux terms are properly known and accounted for in the surface terms of the variational formalism to ensure finite fluxes across the mathematical singularity are allowed and continuity assured. 
\end{itemize}

In view of the very large difference in magnitude between the parallel and perpendicular components of the electric field - itself the consequence of the huge difference in magnitude of the parallel and perpendicular conductivity - the wave equation is 
often expressed 
in terms of $(E_{\perp,1}, E_{\perp,2},E_{//})$ to avoid the (electron) damping involving $E_{//}$ is inaccurately estimated. 
That will also be done here.

In recent years, gradually more powerful - commercial as well as freeware - partial differential equation solvers became available; see e.g. \cite{COMSOL,FENICS,MFEM,FreeFEM}. Grid refinement techniques, higher order polynomial approximations and optimised solvers allow to achieve higher accuracy and/or faster integration. To be able to exploit such tools the reigning equation typically needs to be of (partial) differential form. Finite Larmor radius corrections being important but typically requiring an integro-differential equation, Bud\'{e} proposed a technique to approximate an integro-differential equation by a higher order differential equation \cite{Bude} and found the solutions are hardly distinguisable from the solutions of the actual integro-differential equation. He illustrated its use solving the 1D wave equation corresponding to the "full hot" dielectric tensor of Swanson \cite{Swanson}. Bud\'{e}'s idea was to step away from a Taylor series expansion (which typically breaks down fairly quickly when the adopted "small parameter" $k_\perp\rho_L$ fails to be small
) and adopt a modest order polynomial fit in $\vec{k}$-space. 
From a mathematical point of view, the steps after the fitting are identical to the usual steps: $j$-th powers of $ik_\alpha$ in $\vec{k}$-space become $j$th order partial differential operators. 

The essence of the Bud\'{e}'s procedure proposed goes back to the works of Fuchs et al. \cite{Fuchs-et-al} who - when looking into mode conversion physics - proposed to avoid deriving and solving the full wave equation but just concentrated on the main wave interaction physics: if the evaluation of the dispersion equation $D(x,k)=0$ shows that mode conversion occurs near a reference wave number $\tilde{k}$, then a Taylor series expansion in $\vec{k}$-space yields an approximate dispersion equation 
$$D(x,k) \approx D(x,\tilde{k}) + \frac{dD}{dk}(k-\tilde{k})+\frac{1}{2}\frac{d^2D}{dk^2}(k-\tilde{k})^2=0.$$
near  $k=\tilde{k}$. This expression can immediately be transformed into a second order differential equation by making the Fourier inversion which consists in replacing $k$ by the differential operator $-id/dx$. Although that equation does $not$ fully rigorously describe the whole physics of the 2 interacting modes, it allows to crudely compute how the waves communicate with each other. Bud\'{e} started from this idea to solve the set of 3 coupled wave equations describing the interplay between the 3 electric field components locally exploiting the homogeneous hot plasma dielectric tensor for Maxwellian distributions \cite{Swanson,Stix} and adopting $\tilde{k}_{\perp}=0$ while retaining - as is classically done - only up to second order finite Larmor radius terms. Solving the obtained differential system relying on the finite difference method, he managed to elegantly describe the excitation of short wavelength Bernstein wave branches. He realised he could extend the Taylor series expansion in terms of FLR corrections to \textit{higher order derivatives} without doing much more algebra by numerically \textit{fitting} the tensor locally in $\vec{k}$-space to a higher order polynomial. At the price of augmented \textcolor{black}{efforts needed to obtain the required coefficients of the polynomial fits and of finding the finite difference expressions adopted in the solving scheme he adopted,} 
this permitted him to subsequently solve the wave equation relevant for ion heating at an $arbitrary$ cyclotron harmonic. Solving differential or integro-differential equations relying on the finite element or finite difference technique transforms the initial equation into a linear system yielding the approximate solution. The importance of Bud\'{e}'s work is that he managed to solve the integro-differential equation for all \textit{physically relevant} modes but without needing a \textit{full} system matrix. \textcolor{black}{The subdomain in $\vec{k}$-space where dispersion equation roots appear can be assessed by all-FLR dispersion equation solvers (see e.g. \cite{Stix, KochPHD, VanEester_raytracing_allFLR}). Beyond this domain the Fourier amplitudes are expected to negligibly small or zero and do not contribute to Fourier integrals. The Fourier amplitudes of physically acceptable modes only being nonzero in a finite domain justifies adopting a fit of the relevant functions appearing in the dielectric response provided this fit is sufficiently accurate in the domain of interest.} Since the \textcolor{black}{computer} time and memory required to invert a matrix relying on sophisticated linear system solvers depends sensitively on the amount of nonzero coefficients of the system matrix, this technique potentially allows significantly pushing down the \textcolor{black}{computational} requirements compared to solving the actual integro-differential equation. On top of that, the fit being superior to the Taylor series expansion, Bud\'{e}'s method allows to extend the set of relevant scenarios that can be modelled. His idea was exploited to formulate an extension of the TOMCAT code \cite{TOMCAT} to RF heating scenario modelling requiring an all-FLR integro-differential approach; Maxwellian and bi-Maxwellian distributions can be handled semi-analytically, and - at the price of requiring more \textcolor{black}{computer} time - non-Maxwellian distributions can be treated using a purely numerical approach. The difference between the original (differential) TOMCAT approach and the usual FLR expansion is that it does not make a Taylor series expansion of the dielectric tensor itself but of the Kennel-Engelmann operator \cite{KennelEngelmann1966} acting both on the electric field and on the test function in a variational formulation of the wave equation. Just like is usually done, the original TOMCAT's description relies on a truncation of the dielectric response in terms of FLR corrections at second order but since the operator appears as a product in that response, the equation has up to 4th order Larmor radius corrections. The code's main selling point - and that of its integro-differential as well as Bud\'{e} upgrades - is that it guarantees a positive definite power balance for $any$ of the wave modes when the distribution function is Maxwellian, as is expected from first principles but cannot locally be guaranteed when exploiting the usual expansion for all modes; a fair amount of attention was devoted to studying and settling this issue in the 80's \cite{RomeroScharer,McVey}, the discussion originating from the observation that negative absorption could erroneously occur when a significant amount of the wave energy is carried by short wavelength branches, more specifically by branches that carry their energy via particles in coherent motion with the wave i.e. via the kinetic flux, rather than electromagnetically i.e. via the Poynting flux. Supplementary advantages are that the description allows to go beyond second harmonic heating ($N=2$), and that the wave absorption operator is fully compatible with the quasilinear diffusion operator in the Fokker-Planck equation so a fully self-consistent wave+Fokker-Planck description (in which the $same$ wave-particle interaction description is used) is possible. 

It is tempting to check whether Bud\'{e}'s appealing idea allows to solve the wave equation in 2 or 3 dimensions, generalising the Bud\'{e}-upgraded TOMCAT philosophy to more than 1 dimension. The present paper aims at starting that exercise, \textcolor{black}{providing the basic algebra required and discussing specific points of attention, highlighting both the method's assets and potential limitations/drawbacks}. 

\textcolor{black}{The symmetry of the expression for the dielectric response as exploited in the TOMCAT solver is a simplified "quasi-homogeneous" version of the philosophy originally due to Kaufman \cite{Kaufman}, who demonstrated this symmetry is present on a much deeper level adopting an action-angle approach when assuming the tokamak is axisymmetric. \textcolor{black}{As is also the case in the Kennel-Engelman description, the guiding center rather than the particle position takes central stage in Kaufman's work, removing the need to know the distribution function at the particle position and significantly simplifying the interpretation and algebra.} The orbit is fully determined by 3 constants of the motion along this orbit while the periodic aspects of the motion are described by 3 associated angles: one tracking the Larmor gyration \textcolor{black}{around the guiding centre}, one describing the poloidal bounce motion (the motion ensuring the magnetic moment is conserved when the particle is moving into a region of higher or lower magnetic field strength forcing the $v_\perp$ to change, and - the energy being conserved as well so $v_{//}=0$ can occur - introducing trapping) and one describing the toroidal precession of a toroidal reference point in a toroidal cut. Because of the periodic nature of the particle motion, Fourier analysis is a suitable way of describing the motion. The orbits being closed poloidally, the bounce spectrum consists of a set of discrete modes (labeled as bounce modes) rather than of a continuous spectrum. This has important consequences: constructive and destructive interference isolates the dominant contributions of individual bounce modes to specific poloidal positions when performing the (bounce) integrals on a poloidally closed orbit and allows to reconcile 2 seemingly opposite notions namely (i) the fact that individual bounce modes have global rather than local resonances while (ii) the usual wave-particle interaction relies on the local interaction between wave and particle to explain a net acceleration or deceleration when satisfying the resonance condition $\omega=N\Omega+k_{//}v_{//}$ (or more generally $\omega=N\Omega+\vec{k}.\vec{v}_{D}$ - where $\vec{v}_{D}$ is the guiding centre velocity - when accounting for the deviations of guiding centres from magnetic surfaces). The present paper takes the TOMCAT plasma model "as is" and concentrates on preparing exploitation of Bud\'{e}'s method for numerically solving the integro-differential wave equation in a 2-dimensional space, hereby sidestepping details (such as inhomogeneity corrections resulting from acceleration and deceleration along the orbit) that constitute a very rich research topic in their own right. The expressions provided in the present paper also assume the magnetic surface labeling coordinate and the toroidal angular momentum can be confused. It was illustrated in \cite{AllFLRBudeDVE} that accounting for the actual constants of motion is a technical matter increasing the amount of algebra required but that does not pose particular issues.  
}

The present paper is structured as follows: 
Section \textcolor{black}{2} describes the general starting equation the description relies on. In section \textcolor{black}{3} the general formalism is applied to a Maxwellian distribution and explicit expressions for the needed functions are provided. Section \textcolor{black}{4} generalises these results to bi-Maxwellians with a parallel drift. Section \textcolor{black}{5} briefly mentions how pushing further to arbitrary distributions is possible, at the price of increased \textcolor{black}{computing time}. 
A note on how the general kinetic flux term can be computed is provided in section \textcolor{black}{6}. Section \textcolor{black}{7} is devoted to commenting on a simplified - but limited - description of the parallel dynamics; it also highlights why accounting for the parallel dynamics is challening.  The final form of the wave equation is presented in section \textcolor{black}{8}. 
To allow exploring the potential of the Bud\'{e} method prior to considering all finite Larmor radius corrections, the 2D equivalent of the operator defined in \cite{TOMCAT} is provided in Section \textcolor{black}{9}. It retains up to second order finite Larmor radius corrections in the operator acting both on the electric field and the test function vector and hence yields up to 4th order partial derivatives.
Section \textcolor{black}{10}, finally, sums up the conclusions and comments on the next steps to take towards actual exploitation of the presented expression.

\section{Starting equation}

The original TOMCAT equations were formulated for Maxwellian plasmas. Ichimaru writes the dielectric tensor for an \textit{arbitrary} distribution function in an elegant form \cite{Ichimaru1973}. In view of later generalisation intended, we will initially follow his approach. 
In its most general form, the wave equation for the electric field can be written combining Faraday's and Ampere's law,
\begin{equation}
\nabla \times \nabla \times \vec{E}= i\omega \mu_o \vec{J} +k_o^2\vec{E}=k_o^2 \overline{\overline{\epsilon}}.\vec{E}
\label{eq : Basic} 
\end{equation}
\textcolor{black}{where $\vec{E}$ is the electric field,  $\overline{\overline{\epsilon}}$ is the hot plasma dielectric tensor and $k_o=\omega/c$, where $\omega$ is the driver frequency and $c$ the speed of light.} In the above $\vec{J}$ is the RF perturbed current density 
\begin{equation}
\vec{J}=\sum_{\alpha} q_{\alpha} \int d\vec{v} \vec{v} f_{RF,\alpha}
\end{equation}
in which $f_{RF,\alpha}$ is the RF perturbed distribution function, itself related to the non-perturbed distribution function $F_o$ and to the RF electric and magnetic fields $\vec{E}$ and $\vec{B}$ via Vlasov's equation:
\begin{equation}
f_{RF,\alpha}=-\frac{q_\alpha}{m_\alpha}\int_{-\infty}^{t}dt' [\vec{E}+\vec{v}\times\vec{B}].\nabla_{\vec{v}}F_o.
\end{equation}
Intending later exploitation relying on finite elements, it is suitable to solve the wave equation in variational form. Multiplying the wave equation with the complex conjugate of the test function vector $\vec{F}$, integrating over a region of interest and performing a partial integration on the $\nabla \times \nabla \times \vec{E}$ to bring out the Poynting flux explicitly, the resulting equation can be written 
\begin{equation}
\int d\vec{x} [k_o^2 \vec{F}^*. \overline{\overline{\epsilon}}.\vec{E} -(\nabla \times \vec{F})^*.(\nabla \times \vec{E})]=-\int_S d\vec{S}.\vec{F}^*\times \nabla \times \vec{E}.
\label{eq : BasicVar} 
\end{equation}
\begin{figure}   [ht!]
 \includegraphics[width=4.in]{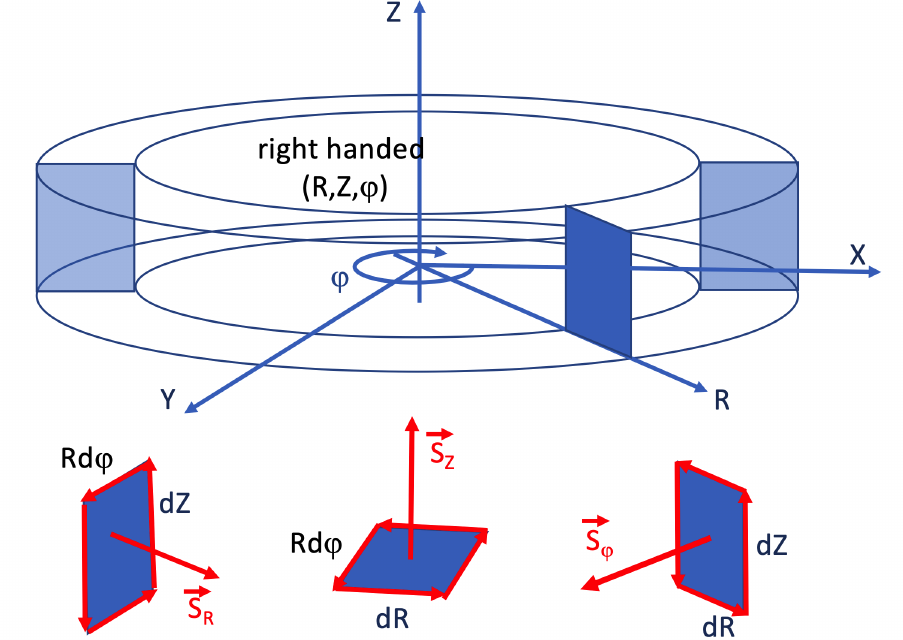}
\caption{Right handed cylindrical $(R,Z,\varphi)$ coordinate system and elementary surfaces. } 
 \label{figure : CYL} 
\end{figure}
Following Ichimaru and adopting cylindrical coordinates $(R,Z,\varphi)$ - see Fig. \ref{figure : CYL} - the term involving the dielectric tensor can be written as
\begin{equation}
\begin{split}
\vec{F}^*.\overline{\overline{\epsilon}}.\vec{E}=[1-\sum_\alpha \frac{\omega_p^2}{\omega^2}] \vec{F}^*.\overline{\overline{1}}.\vec{E} - 2\pi \sum_\alpha \frac{\omega_p^2}{\omega^2} \sum_{N=-\infty}^{+\infty} \int_{0}^{\infty} dv_\perp \int_{-\infty}^{+\infty} dv_{//} \int dk'_R dk'_Z \int dk_R dk_Z   \\
\sum_{n,n'} e^{i([k_R-k'_R]R+[k_Z-k'_Z]Z+[n-n']\varphi)}\frac{[ N \Omega_{\alpha} \partial F_o/ \partial v_{\perp} +k_{//}v_{\perp} \partial F_o/ \partial v_{//} ]}{ [ N \Omega_\alpha + k_{//}v_{//}-\omega]} L(\vec{F}_{\vec{k}'})^*L(\vec{E}_{\vec{k}})
\end{split}
\label{eq : Ichi} 
\end{equation}
where $n$ and $n'$ are the toroidal mode numbers of $\vec{E}$ and $\vec{F}$, respectively; $\vec{F}$ is the test function vector, $\vec{E}$ is the electric field and $L$ is the Kennel-Engelmann operator \cite{KennelEngelmann1966} 
\begin{equation}
L(\vec{H})=\frac{v_\perp}{2} \Big [H_{-}J_{N+1}exp[i(N+1)\psi]+H_{+}J_{N-1}exp[i(N-1)\psi] \Big ] +v_{//} H_{//}J_{N}exp[iN\psi] 
\end{equation}
in which $ H\pm=H_{\perp,1}\pm iH_{\perp,2}$ and $ \psi =tan^{-1}( k_{\perp,2}/k_{\perp,1} )$; the argument of the Bessel functions is $\zeta=k_{\perp}\rho_L=k_\perp v_\perp/\Omega_\alpha$ and we will henceforth assume the tokamak is axisymmetrical so that the various toroidal modes can be treated one by one as finite contributions require $n=n'$. Note that formally the treatment can easily be extended to 3D by retaining the full sum on $n$ and $n'$ and hence treating all toroidal couplings when projecting the ultimately obtained equation on a full set of toroidal mode test functions rather than just one. 
The argument $\zeta$ contains $k_\perp= (k_{\perp,1}^2+k_{\perp,2}^2)^{1/2}$. Via the angle $\psi$ the dielectric tensor accounts for directionality of the wave. 

It has been assumed that the slowly varying distribution function does not depend on the cyclotron gyro-phase. More generally and thinking beyond 1D application, the slowly varying distribution $F_o$ does not vary on any of the 3 oscillatory fastly varying aspects of the motion: cyclotron gyro-motion, poloidal bounce motion or toroidal precession drift motion. 
We put the expression in matrix form for easy manipulation.
$$\vec{F}^*.\overline{\overline{\epsilon}}.\vec{E}=[1-\sum_\alpha \frac{\omega_p^2}{\omega^2}] \vec{F}^*.\overline{\overline{1}}.\vec{E} - 2\pi \sum_\alpha \frac{\omega_p^2}{\omega^2} \sum_{N=-\infty}^{+\infty} \int_{0}^{\infty} dv_\perp \int_{-\infty}^{+\infty} dv_{//} \int dk'_R dk'_Z \int dk_R dk_Z  $$
$$e^{i([k_R-k'_R]R+[k_Z-k'_Z]Z+[n-n']\varphi)}\frac{[ N \Omega_{\alpha} \partial F_o/ \partial v_{\perp} +k_{//}v_{\perp} \partial F_o/ \partial v_{//} ]}{ [ N \Omega_\alpha + k_{//}v_{//}-\omega]} $$
\begin{equation}
\Bigg ( \begin{array}{ccc} F_{\perp,1} & F_{\perp,2} & F_{//} \end{array} \Bigg )_{\vec{k}'}^* . \textcolor{black}{\overline{\overline{\textrm{M}}}_{diel}}.  \Bigg ( \begin{array}{c} E_{\perp,1} \\ E_{\perp,2} \\ E_{//} \end{array} \Bigg )_{\vec{k}} 
\end{equation}
Here 
\textcolor{black}{
\begin{equation}
\overline{\overline{\textrm{M}}}_{diel}=\Bigg ( \begin{array}{ccc} G_{N,1}(\vec{k}) G_{N,1}^*(\vec{k}')&G_{N,2}(\vec{k})  G_{N,1}^*(\vec{k}')& G_{N,3}(\vec{k}) G_{N,1}^*(\vec{k}') \\   G_{N,1} G_{N,2}^*(\vec{k}') (\vec{k}) &G_{N,2}(\vec{k}) G_{N,2}^*(\vec{k}') & G_{N,3}(\vec{k})  G_{N,2}^*(\vec{k}') \\ G_{N,1}(\vec{k}) G_{N,3}^*(\vec{k}') &G_{N,2}(\vec{k}) G_{N,3}^*(\vec{k}') & G_{N,3}(\vec{k})  G_{N,3}^*(\vec{k}') \end{array} \Bigg )
\end{equation}
}
\textcolor{black}{and}
$$G_{N,1}(\vec{k})=\frac{v_\perp}{2}[+J_{N+1}e^{i\psi}+J_{N-1}e^{-i\psi}]e^{iN\psi}$$
$$G_{N,2}(\vec{k})=\frac{iv_\perp}{2}[-J_{N+1}e^{i\psi}+J_{N-1}e^{-i\psi}]e^{iN\psi}$$
\begin{equation}
G_{N,3}(\vec{k})=v_{//}J_N e^{iN \psi} 
\end{equation}

\textcolor{black}{It is important to underline that} the dielectric operator $\overline{\overline{\epsilon}}$ in Eq.\ref{eq : Basic} and the above are actually \textit{not} identical: In the symmetrical weak variational principle formulation, $\overline{\overline{\epsilon}}$ not only operates to the right on the electric field $\vec{E}$ but also to the left on the test function $\vec{F}$, while $\vec{F}$ does not even appear in Eq.\ref{eq : Basic} so that the operator can only act on $\vec{E}$. 
\textcolor{black}{For that reason and despite the fact that the Kennel-Engelman operator was derived for a homogeneous plasma, the above expression is not just a generalisation of the uniform plasma expression, the distinction lying in the fact that the present expression for the dielectric response has one Kennel-Engelman operator acting on the test function (so that $\vec{k}'$ appears in it)  while the other is acting on the electric field (where $\vec{k}$ appears). The uniform plasma expression would have both operators having $\vec{k}$ so that the test function can simply be moved in front (as in the first equation) while that is no longer possible in the above. Unlike what is expected from first principles  when the populations are in thermal equilibrium and as noticed early on by various authors, assembling a wave equation by adopting the homogeneous plasma limit and merely substituting $\vec{k}$'s by $-i\nabla$'s does \textit{not} guarantee that energy exchange between particles and waves always flows from the waves to the particles. The proper placing of the differential operators in the expression dates back from a discussion in the 80's on the meaning of what we understand under "heating" of a population that is confined by a strong magnetic field \cite{RomeroScharer,McVey} and yields a dielectric tensor operator that contains derivatives w.r.t. background quantities. The adopted model retains leading order terms in $\rho_L/L$ (Larmor radius on equilibrium scalelength) while ensuring positive definiteness of any wave mode the plasma admits but sidestepping deriving basic equations to account for supplementary - less crucial - corrections due to the inhomogeneity of the background.} 

The choice of the grid is a matter of discussion. The traditional procedure is to use a grid aligned with the magnetic surfaces, which is also convenient for treating the wall (which can then be treated as a coordinate surface when details of the vessel's shape are neglected). This suggests to use the poloidal angle as one of the variables and would allow Fourier analysis in that direction, which has the major advantage that $k_{//}$ remains a well defined \textit{algebraic} quantity and allows to keep using the plasma dispersion function when introducing the poloidal field effects. It has the drawback that \textcolor{black}{there is a singularity of the coordinate system at the magnetic axis}, although physically the axis is a perfectly regular point. The other option is to use a "Cartesian" $(R,Z)$ grid in a toroidal cut. This allows taking over almost all of the machinery developed for the 1D application - up to upgrading the \textcolor{black}{algebra to permit 2D exploitation} - but has the drawback that the wall is not a coordinate surface. It is - however - perfectly feasible to have a complicated wall structure when imposing the wall boundary conditions via Lagrange multipliers \cite{LagrangeMultipliers}. An intermediate procedure seems most appropriate: adopting a "Cartesian" $(R,Z)$ grid in a toroidal cut but exploiting triangles as elementary finite elements in which the wave equation is solved exploiting a set of base functions of sufficiently high order for the equation at hand. Powerful finite element packages are available allowing exactly that; e.g. MFEM \cite{MFEM} supports a wide variety of finite element spaces in 2D and 3D, including arbitrary high-order representation. Exploiting them, wall or plasma edge surfaces that are not coordinate surfaces can be treated elegantly while all finite element \textcolor{black}{algebra to construct and exploit the proper base functions} is done internally in MFEM routines. 

\begin{figure}   [ht!]
 \includegraphics[width=4.in]{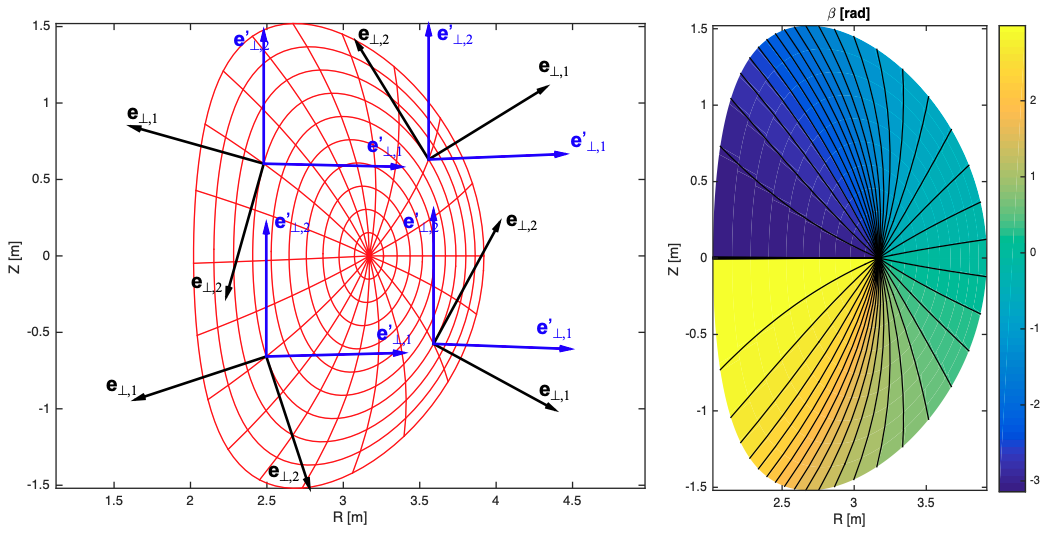}
\caption{Most commonly exploited $(\vec{e}_{\perp,1},\vec{e}_{\perp,2})=(\vec{e}_\rho,\vec{e}_\theta)$ and modified $(\vec{e'}_{\perp,1},\vec{e'}_{\perp,2})$ perpendicular unit vectors (left) and angle $\beta$ (right)\textcolor{black}{.} }
 \label{figure : FRAME} 
\end{figure}

When the poloidal magnetic field is accounted for, the \textcolor{black}{mathematical and numerical effort needed is increased.} 
A rotation matrix $\overline{\overline{\mathcal{R}}}$ now connects the \textcolor{black}{locally adopted $(\vec{e'}_{\perp,1},\vec{e'}_{\perp,2},\vec{e}_{//})$} and the global or geometrical $(\vec{e}_R,\vec{e}_Z,\vec{e}_\varphi)$ frames. Although for the perpendicular direction the procedure is essentially unchanged, the parallel direction requires more attention since $k_{//}$ now has a poloidal component. Labeling $\Theta$ to be the angle between the toroidal and the parallel direction one has
\begin{equation}
k_{//}=\textcolor{black}{\cos} \Theta \frac{n}{R} + \textcolor{black}{\sin} \Theta k_{\theta}=\textcolor{black}{\cos} \Theta k_{\varphi} + \textcolor{black}{\sin} \Theta k_{\theta} 
\end{equation}
in which the poloidal wave vector component $k_\theta$ itself is a function of $k_R$ and $k_Z$ in general. \textcolor{black}{The} 2 perpendicular directions \textcolor{black}{defined via the unit vectors  $\vec{e'}_{\perp,1}$  and  $\vec{e'}_{\perp,2}$ } can be defined to be "as close as possible" to $\vec{e}_R$ and $\vec{e}_Z$\textcolor{black}{;}
see Fig. \ref{figure : FRAME}. \textcolor{black}{
This requires 
\begin{equation}
\frac{d}{d\beta} [\vec{e'}_{\perp,1} . \vec{e}_R]=\frac{d}{d\beta} [\textcolor{black}{\cos} \beta \vec{e}_{\perp,1} . \vec{e}_R+\textcolor{black}{\sin} \beta \vec{e}_{\perp,2} . \vec{e}_R]=0 
\end{equation}
\textcolor{black}{where $\vec{e}_{\perp,1}=\nabla \rho / |\nabla \rho|$ and $\vec{e}_{\perp,2}=[\partial \theta/\partial \vec{x}] / |\partial \theta/\partial \vec{x}|$ (with $\rho$ the magnetic surface labeling parameter and $\theta$ the poloidal angle) so} 
$$\textcolor{black}{\sin} \beta =\frac{-\textcolor{black}{\cos} \Theta \textcolor{black}{\sin} \alpha}{[\textcolor{black}{\cos}^2\alpha +(\textcolor{black}{\cos}\Theta \textcolor{black}{\sin} \alpha)^2 ]^{1/2}},$$
\begin{equation}
\textcolor{black}{\cos} \beta =\frac{\textcolor{black}{\cos} \alpha}{[\textcolor{black}{\cos}^2\alpha +(\textcolor{black}{\cos}\Theta \textcolor{black}{\sin} \alpha)^2 ]^{1/2}}. 
\end{equation}
in which $\alpha$ is the angle between $\vec{e}_R$ and $\vec{e}_{\perp,1}$. This elegant procedure was already exploited in AORSA \cite{AORSA}.}

There is a supplementary subtlety: $k_{//}$ appears in the factor introducing the resonant denominator. Depending on the approximations used, $k_{//}=k_{//}(\vec{k})$ or $k_{//}=k_{//}(\vec{k},\vec{k}')$. The former is most frequently used in modelling while the latter is aligned with the leading order contribution picked up when accounting for the bounce dynamics but subsequently omitting the sum on the bounce modes by only retaining the dominant contribution of the relevant discrete bounce sum transformed into the corresponding bounce integral (in that case the relevant $k_{//}$ to use in the resonant denominator is $k_{//}=\textcolor{black}{\cos} \Theta n/R + \textcolor{black}{\sin} \Theta (k_\theta+k'_\theta)/2$; see e.g \cite{LamalleBIGpaper,VanEesterPolBounce,VanEesterPolBounce2}).
Aligned with more general principles (formulated in the visionary paper by Kaufman \cite{Kaufman}), the latter approach guarantees a positive definite power absorption for $any$ wave a plasma composed of species in thermal equilibrium  (i.e. having Maxwellian distribution functions) admits. 

\textcolor{black}{
We will henceforth adopt the notation
}
\textcolor{black}{
\begin{equation}
\overline{\overline{\mathcal{P}}}_0=[1-\sum_\alpha \frac{\omega_p^2}{\omega^2}] \overline{\overline{1}} 
\end{equation}
}
\textcolor{black}{
and will label the remainder of the expression for the operator $\overline{\overline{\epsilon}}$ as $\overline{\overline{\mathcal{P}}}_1$
}
in which the latter - following the procedure proposed by Bud\'{e} - is subsequently fitted using a multipolynomial fit 
\begin{equation}
\overline{\overline{\mathcal{P}}}_1= \sum_{I,J,K,L}   \overline{\overline{\mathcal{P}}}_{1,I,J,K,L} k_R^I  k_Z^J  (k'_R)^K  (k'_Z)^L 
\end{equation}
where all sums are truncated at a predetermined order so that the dielectric response term in the wave equation can be written as a sum of partial differential equation contributions 
\begin{equation}
\vec{F}^*.\overline{\overline{\epsilon}}.\vec{E}=\vec{F}^*.\overline{\overline{\mathcal{P}}}_0.\vec{E}-\sum_{I,J,K,L}  (-i)^{I+J-K-L} \frac{\partial^{K+L} \vec{F}^* }{\partial R^K \partial Z^L}.\overline{\overline{\mathcal{P}}}_{1,I,J,K,L}.\frac{\partial^{I+J}\vec{E}}{\partial R^I \partial Z^J} 
\end{equation}
in which the vectors are expressed in terms of their cylindrical components. The Fourier integrations have been done relying on 
\begin{equation}
\frac{\partial^{I+J}H}{\partial R^I \partial Z^J}(\vec{x}) = (+i)^{I+J} \int dk_R dk_Z exp[i(k_R R + k_Z Z +n\varphi)] k_R^I k_Z^J H_{\vec{k}}. 
\end{equation}
based on the Fourier representation
\begin{equation}
H(\vec{x}) = \int dk_R dk_Z exp[i(k_R R + k_Z Z +n\varphi)] H_{\vec{k}}. 
\end{equation}
In the particular case that the poloidal field is zero, the computation is much simplified: In that case $k_{//}$ \textit{only} depends on $n$, $R$ and $Z$, 
\textcolor{black}{yielding expressions of the form}
\begin{equation}
\sum_{N=-\infty}^{\infty} \frac{[ N \Omega_{\alpha} \partial F_o/ \partial v_{\perp} +k_{//}v_{\perp} \partial F_o/ \partial v_{//} ]}{ [ N \Omega_\alpha + k_{//}v_{//}-\omega]}  G_{N,i}(\vec{k}_{\perp}(k_R,k_Z))G_{N,j}^*(\vec{k}'_{\perp}(k'_R,k'_Z)) 
\end{equation}
\textcolor{black}{in which $G_{N,i}(\vec{k}_\perp)$ and $G_{N,j}(\vec{k}_\perp')$ can be fitted separately i.e. in which only 2D fits in terms of  either $(k_R,k_Z)$  or $(k_R',k_Z')$ - rather than 4D fits of products of $G_{N,...}$ and including the resonant denominator in terms of both $(k_R,k_Z)$ and $(k_R',k_Z')$ - need to be made. } 

\section{Derivation of practical expressions for Maxwellian $F_o$}

In case the distributions are Maxwellian i.e. 
\begin{equation}
F_o=\frac{1}{[2\pi]^{3/2}v_t^3}exp[-\frac{(v_\perp^2+v_{//}^2)}{[2v_t^2]}] 
\end{equation}
then the expression for $\overline{\overline{\mathcal{P}}}_1$ becomes
\begin{equation} \label{EQ:termmax}
\overline{\overline{\mathcal{P}}}_1= -2\pi \sum_\alpha \frac{\omega_p^2}{\omega^2} \sum_{N=-\infty}^{+\infty} \int_{0}^{\infty} dv_\perp \int_{-\infty}^{+\infty} dv_{//}   \frac{v_\perp F_o}{v_t^2}\Big [ 1+\frac{\omega}{[ N \Omega_\alpha + k_{//}v_{//}-\omega]} \Big ] \textcolor{black}{\overline{\overline{\textrm{M}}}_{diel}} 
\end{equation}
The following expression for evaluating the integrals of Bessel function products is handy \cite{Stix}:
\begin{equation}
\mathcal{I}_1(\nu,a,b,p)=\int_0^\infty t J_\nu(at)J_\nu(bt)e^{-p^2t^2}dt=\frac{1}{2p^2}exp[-\frac{a^2+b^2}{4p^2}]I_\nu(\frac{ab}{2p^2})  
\end{equation}
which holds for $Re[\nu]>-1$ and $|arg(p)|<\pi/4$. For the latter, no action is required since $p$ is real for our application. The former requires exploiting the relations of the Bessel functions to obtain the suitable expressions. The identity $J_{-N}=(-)^NJ_N$ readily states that it suffices to use $|N|$ when evaluating the expressions, as also follows from $I_{-N}=I_N$. Taking the derivative w.r.t. $a$ yields
\begin{equation}
\mathcal{I}_2(\nu,a,b,p)=\int_0^\infty t^2 J'_\nu(at)J_\nu(bt)e^{-p^2t^2}dt=\frac{1}{4p^4}exp[-\frac{a^2+b^2}{4p^2}][-a I_\nu(\frac{ab}{2p^2})+bI'_\nu(\frac{ab}{2p^2}) ]  
\end{equation}
and taking the derivative w.r.t. both $a $ and $b$ yields
\begin{equation}
 \mathcal{I}_3(\nu,a,b,p)=\int_0^\infty t^3 J'_\nu(at)J'_\nu(bt)e^{-p^2t^2}dt=\frac{1}{4p^4}exp[-\frac{a^2+b^2}{4p^2}] [\frac{ab}{2p^2}(I_N+I_N'')+(1-\frac{(a^2+b^2)}{2p^2})I'_N]    
\end{equation}
in which the quotes are the derivatives w.r.t. the argument of the respective functions. The following expressions are useful:
$$I'_\nu=\frac{1}{2}[I_{\nu+1}+I_{\nu-1}]$$
\begin{equation}
I''_{\nu}(z)=\frac{[z^2+\nu^2]I_\nu -z I'_\nu}{z^2}  
\end{equation}
It was checked both analytically and numerically that the latter does not diverge at $z=0$. \textcolor{black}{ The finite $\nu^2 I_\nu$ term compensates $z I'_\nu$ term so that the denominator going to zero does not cause a problem.} For evaluating these functions \textit{numerically} it is needed to evaluate the expressions slightly away from $z=0$ to avoid overflows. 

We write the matrix in terms of the Bessel functions and their derivatives.
which yields the specific values $\nu=N$, $t=v_\perp$, $a=k_\perp/\Omega$, $b=k'_\perp/\Omega$ and $p=1/[2^{1/2}v_t]$. The required parallel integrals are of the form
\begin{equation}
I_{//,j}(\xi)=\int_{-\infty}^{+\infty}dt \frac{t^j}{t-\xi} e^{-t^2} 
\end{equation}
where 
\begin{equation}
I_{//,0}(\xi)=\pi^{1/2} \mathcal{Z}(\xi)  
\end{equation}
with $\mathcal{Z}$ the plasma dispersion function \cite{FriedConte}. The recursion
\begin{equation}
I_{//,M}=\tilde{I}_{M-1}+\xi I_{//,M-1} 
\end{equation}
where
\begin{equation}
\tilde{I}_M=\int_{-\infty}^{+\infty}dt e^{-t^2}t^M 
\end{equation}
allows to find the needed expressions. The integrals $\tilde{I}_M$ themselves are zero for odd $M$ and can be found from the recursion
\begin{equation}
\tilde{I}_M=\frac{[M-1]}{2}\tilde{I}_{M-2} 
\end{equation}
for even $M$. Starting from $\tilde{I}_0=\pi^{1/2}$ we get $\tilde{I}_2=\pi^{1/2}/2$. Then $I_{//,1}=\tilde{I}_0+\xi I_{//,0}=\pi^{1/2}[1+\xi \mathcal{Z}(\xi)]$ and $I_{//,2}=\tilde{I}_1+\xi I_{//,1}=\pi^{1/2} \xi (1+\xi \mathcal{Z}(\xi))$. We can now perform all velocity space integrals. The result for contributions to $\textcolor{black}{-}\mathcal{P}_1$ and hence directly to be summed to the $[1-\omega_p^2/\omega^2] \overline{\overline{1}}$ term is   
$$2\pi \sum_\alpha \frac{\omega_p^2}{\omega^2} \sum_{N=-\infty}^{+\infty} \int_{0}^{\infty} dv_\perp  \int_{-\infty}^{+\infty} dv_{//}   \frac{v_\perp F_o}{v_t^2}\Big [ 1+\frac{\omega}{[ N \Omega_\alpha + k_{//}v_{//}-\omega]} \Big ] G_{N,1}(\vec{k}) G_{N,1}^*(\vec{k}')=$$
$$ \sum_\alpha \frac{\omega_p^2}{\omega^2v_t^4} \sum_{N=-\infty}^{+\infty}  \Big [1+\frac{\omega}{k_{//}2^{1/2}v_t} \mathcal{Z}_N \Big ] $$
$$[\frac{(N\Omega)^2}{k_\perp k'_\perp}\textcolor{black}{\cos} \psi \textcolor{black}{\cos} \psi' \mathcal{I}_1(\nu,a,b,p)+\frac{iN\Omega }{k_\perp}\textcolor{black}{\cos}\psi \textcolor{black}{\sin}\psi' \mathcal{I}_2(\nu,b,a,p)   $$
\begin{equation}-\frac{iN\Omega}{k'_\perp}\textcolor{black}{\sin}\psi \textcolor{black}{\cos}\psi' \mathcal{I}_2(\nu,a,b,p)+ \textcolor{black}{\sin}\psi \textcolor{black}{\sin}\psi' \mathcal{I}_3(\nu,a,b,p)  ]   \Big ] e^{iN[\psi-\psi']} 
\end{equation}
where $\xi_N=[\omega-N\Omega_\alpha]/[k_{//}2^{1/2}v_t]$. 
\textcolor{black}{Introducing the notation
$$\textbf{\textit{R}}_{\alpha,\beta}=2\pi \sum_\alpha \frac{\omega_p^2}{\omega^2} \sum_{N=-\infty}^{+\infty} \int_{0}^{\infty} dv_\perp  \int_{-\infty}^{+\infty} dv_{//}   \frac{v_\perp F_o}{v_t^2}\Big [ 1+\frac{\omega}{[ N \Omega_\alpha + k_{//}v_{//}-\omega]} \Big ] G_{N,\alpha}(\vec{k}) G_{N,\beta}^*(\vec{k}')$$
}
\textcolor{black}{so that $\overline{\overline{\textbf{\textit{R}}}}$ is the transpose of $\overline{\overline{\textbf{\textit{P}}}}_1$}
\textcolor{black}{we similarly get}
$$\textcolor{black}{\textbf{\textit{R}}_{1,2}} = \sum_\alpha \frac{\omega_p^2}{\omega^2v_t^4} \sum_{N=-\infty}^{+\infty}  \Big [ 1+\frac{\omega}{k_{//}2^{1/2}v_t}\mathcal{Z}_N \Big ] 
$$
$$
\Big [ -\frac{i\Omega N }{k_\perp}\textcolor{black}{\cos}\psi \textcolor{black}{\cos}\psi' \mathcal{I}_2(\nu,b,a,p)\textcolor{black}{+}\frac{(N\Omega)^2}{k_\perp k'_\perp}\textcolor{black}{\cos}\psi \textcolor{black}{\sin} \psi' \mathcal{I}_1(\nu,a,b,p)
$$
$$\textcolor{black}{-}\mathcal{I}_3(\nu,a,b,p)\textcolor{black}{\sin}\psi \textcolor{black}{\cos} \psi' - \frac{i N \Omega }{k'_\perp} \textcolor{black}{\sin}\psi \textcolor{black}{\sin}\psi' \mathcal{I}_2(\nu,a,b,p) \Big ] e^{iN[\psi-\psi']} $$
$$\textcolor{black}{\textbf{\textit{R}}_{1,3}}= \sum_\alpha \frac{\omega_p^2}{\omega^2v_t^4} \sum_{N=-\infty}^{+\infty}  \Big [ \frac{\omega}{k_{//}} [1+\xi_N \mathcal{Z}_N] \Big ] $$
$$[\frac{N\Omega }{k_\perp}\textcolor{black}{\cos}\psi  \mathcal{I}_1(\nu,a,b,p) \textcolor{black}{-}i \mathcal{I}_2(\nu,a,b,p) \textcolor{black}{\sin} \psi]  \Big ] e^{iN[\psi-\psi']}$$
$$\textcolor{black}{\textbf{\textit{R}}_{2,1}}= \sum_\alpha \frac{\omega_p^2}{\omega^2v_t^4} \sum_{N=-\infty}^{+\infty}  \Big [1+\frac{\omega}{k_{//}2^{1/2}v_t}\mathcal{Z}_N\Big ] $$
$$ \Big [ \frac{iN\Omega }{k'_\perp}\textcolor{black}{\cos}\psi \textcolor{black}{\cos} \psi' \mathcal{I}_2(\nu,a,b,p)\textcolor{black}{+}\frac{(N\Omega)^2}{k_\perp k'_\perp} \textcolor{black}{\sin} \psi \textcolor{black}{\cos} \psi' \mathcal{I}_1(\nu,a,b,p) $$
$$\textcolor{black}{-} \textcolor{black}{\cos}\psi \textcolor{black}{\sin}\psi' \mathcal{I}_3(\nu,a,b,p)+\frac{iN\Omega}{k_\perp} \textcolor{black}{\sin}\psi \textcolor{black}{\sin} \psi' \mathcal{I}_2(\nu,b,a,p) \Big  ]  e^{iN[\psi-\psi']}$$
$$\textcolor{black}{\textbf{\textit{R}}_{2,2} }= \sum_\alpha \frac{\omega_p^2}{\omega^2v_t^4} \sum_{N=-\infty}^{+\infty}  \Big [ 1+\frac{\omega}{k_{//}2^{1/2}v_t}\mathcal{Z}_N \Big ] $$
$$\Big [ \mathcal{I}_3(\nu,a,b,p) \textcolor{black}{\cos} \psi \textcolor{black}{\cos} \psi' \textcolor{black}{-}\frac{iN \Omega }{k_\perp} \mathcal{I}_2(\nu,b,a,p)\textcolor{black}{\sin}\psi \textcolor{black}{\cos} \psi'$$
$$\textcolor{black}{+}\frac{iN\Omega }{k_\perp'}\textcolor{black}{\cos} \psi \textcolor{black}{\sin} \psi' \mathcal{I}_2(\nu,a,b,p)+\frac{(N\Omega)^2}{k_\perp k'_\perp}\textcolor{black}{\sin} \psi \textcolor{black}{\sin} \psi' \mathcal{I}_1(\nu,a,b,p) \Big ]   e^{iN[\psi-\psi']}$$
$$\textcolor{black}{\textbf{\textit{R}}_{2,3}}= \sum_\alpha \frac{\omega_p^2}{\omega^2v_t^4} \sum_{N=-\infty}^{+\infty} \Big [ \frac{\omega}{k_{//}}[1+\xi_N \mathcal{Z}_N] \Big ] \Big [ i  \mathcal{I}_2(\nu,a,b,p) \textcolor{black}{\cos} \psi+\frac{N\Omega }{k_\perp}\mathcal{I}_1(\nu,a,b,p)\textcolor{black}{\sin} \psi \Big ] e^{iN[\psi-\psi']} $$
$$\textcolor{black}{\textbf{\textit{R}}_{3,1}} = \sum_\alpha \frac{\omega_p^2}{\omega^2v_t^4} \sum_{N=-\infty}^{+\infty} \Big [ \frac{\omega}{k_{//}} [1+\xi_N \mathcal{Z}_N] \Big ] \Big [ \frac{N\Omega }{k'_\perp} \textcolor{black}{\cos} \psi' \mathcal{I}_1(\nu,a,b,p)\textcolor{black}{+}i\mathcal{I}_2(\nu,b,a,p) \textcolor{black}{\sin} \psi' \Big ]  e^{iN[\psi-\psi']}$$
$$\textcolor{black}{\textbf{\textit{R}}_{3,2}} = \sum_\alpha \frac{\omega_p^2}{\omega^2v_t^4} \sum_{N=-\infty}^{+\infty} \Big [ \frac{\omega}{k_{//}} [1+\xi_N \mathcal{Z}_N] \Big ] \Big [ -i  \mathcal{I}_2(\nu,b,a,p) \textcolor{black}{\cos} \psi' \textcolor{black}{+}\frac{N\Omega }{k'_\perp} \mathcal{I}_1(\nu,b,a,p) \textcolor{black}{\sin} \psi' \Big ]  e^{iN[\psi-\psi']} $$
\begin{equation} \label{EQ:1Dfinal}
\textcolor{black}{\textbf{\textit{R}}_{3,3}} = \sum_\alpha \frac{\omega_p^2}{\omega^2v_t^2} \sum_{N=-\infty}^{+\infty}   \Big [ 1+\frac{2\omega}{k_{//}2^{1/2}v_t}\xi_N [1+\xi_N \mathcal{Z}_N ] \Big ]  \mathcal{I}_1(\nu,b,a,p)  e^{iN[\psi-\psi']}  
\end{equation}

\textcolor{black}{Realising that the leading order term in the asymptotic expansion of both $I_N$ and its derivative is $exp[z]/[2\pi z]^{1/2}$, the exponential terms in the integrals combine to yield $exp[-(a - b)^2/(2p)^2] = exp[-(k_\perp - k'_\perp )^2\rho^2_L/2]$ showing that contributions are small when $k_\perp$ and $k'_\perp$ differ significantly.}

\section{Upgrade to bi-Maxwellian distributions with parallel drift}

In case the distribution is a bi-Maxwellian with perpendicular temperature $T_\perp$, parallel temperature $T_{//}$ and parallel velocity drift $v_{//,d}$ so that 
\begin{equation}
F_o=\frac{1}{[2\pi]^{3/2}v_{t,\perp}^2v_{t,//}}exp[-\frac{v_\perp^2}{2v_{t,\perp}^2}] exp[-\frac{(v_{//}-v_{//,d})^2}{2v_{t,//}^2}]  
\end{equation}
where $v_{t,\perp}=[kT_{\perp}/m]^{1/2}$ and $v_{t,//}=[kT_{//}/m]^{1/2}$, the term
\begin{equation}
\frac{v_\perp F_o}{v_t^2}\Big [ 1+\frac{\omega}{[ N \Omega_\alpha + k_{//}v_{//}-\omega]} \Big ]
\end{equation}
in Eq. \ref{EQ:termmax} needs to be upgraded to
\begin{equation}
\frac{v_\perp F_o}{v_{t,//}^2}\Big [ 1+\frac{\omega+N\Omega_\alpha[(v_{t,//}/v_{t,\perp})^2-1]-k_{//}v_{//,d}}{[ N \Omega_\alpha + k_{//}v_{//}-\omega]} \Big ].
\end{equation}
Since - aside from the upgraded $F_o$ - no new dependence on $v_\perp$ or $v_{//}$ is introduced but only the coefficients in the above differ from the already obtained results, the upgrade to this more general type of distribution merely requires to replace $v_t$ by $v_{t,\perp}$ in the perpendicular integrals (a factor $1/v_{t,\perp}^2$ already appearing from the expression of $F_o$ and further contributions related to the $\mathcal{I}_{...}$ integrals), $v_t$ by $v_{t,//}$ in the parallel integrals, to upgrade the argument of the 
\textcolor{black}{plasma dispersion} function $\xi_{N}=[\omega-N\Omega]/[k_{//}2^{1/2}v_{t,//}]$ to $\xi_{N,d}=[\omega-N\Omega-k_{//}v_{//,d}]/[k_{//}2^{1/2}v_{t,//}]$ generalising the notation $\mathcal{Z}_{N}$ to $\mathcal{Z}_{N,d}$ and adjusting the elementary parallel integrals to account for the extra drift while upgrading the proper coefficients $1$ for the nonresonant and $\omega$ for the resonant contributions in the straight brackets together with the $1/v_{t}^2$ in front of it in Eq.\ref{EQ:1Dfinal} to $1/v_{t,//}^2$ and $[\omega+N\Omega_\alpha[(v_{t,//}/v_{t,\perp})^2-1]-k_{//}v_{//,d}]/v_{t,//}^2$. While the perpendicular integrals are untouched except for introducing the proper thermal velocity, the parallel ingredients are mildly upgraded using the splitting $v_{//}=[v_{//}-v_{//,d}]+v_{//,d}$ and relying on partial integration. Labeling 
$I_{NR //,d,m}=\int dv_{//} v_{//}^m exp[-\frac{(v_{//}-v_{//,d})^2}{2v_{t,//}^2}]$
\begin{equation}
=(2^{1/2}v_{t,//})^{m+1}\int dq (q+q_d)^m exp[-q^2]=(2^{1/2}v_{t,//})^{m+1}I_{NR,m}
\end{equation}
we can easily find that 
\begin{equation}
I_{NR,m}=\frac{m-1}{2}I_{NR,m-2}+q_dI_{NR,m-1}
\end{equation}
so that we get the 3 needed nonresonant parallel integrals $I_{NR//,d,0}=2^{1/2}v_{t,//}\pi^{1/2}$, $I_{NR//,d,1}=2v_{t,//}^2q_d\pi^{1/2}$ and $I_{NR//,d,2}=2^{3/2}v_{t,//}^3\pi^{1/2}[q_d^2+1/2]$ where $q_d=v_{//,d}/[2^{1/2}v_{t,//}]$. For the resonant integrals and labeling 
$$
I_{R//,d,m}=\int dv_{//} \frac{v_{//}^m}{[N \Omega_\alpha + k_{//}v_{//}-\omega]} exp[-\frac{(v_{//}-v_{//,d})^2}{2v_{t,//}^2}]
$$
\begin{equation}
=\frac{(2^{1/2}v_{t,//})^m}{k_{//}}\int dq \frac{(q+q_d)^m}{[q-\xi_{N,d}]} exp[-q^2]=\frac{(2^{1/2}v_{t,//})^m}{k_{//}}I_{R,m}
\end{equation}
for which the recursion 
\begin{equation}
I_{R//,d,m}=I_{NR//,d,m-1}+(\xi_{N,d}+q_d)I_{R//,d,m-1}
\end{equation}
holds, we get $I_{R//,d,0}=\pi^{1/2}\mathcal{Z}_{0,d}/k_{//}$, 
$I_{R//,d,1}=[2^{1/2}v_{t,//}/k_{//}]\pi^{1/2}[1+\xi_{1,d,0}\mathcal{Z}_{1,d}]$ and 
$I_{R//,d,2}=[2v_{t,//}^2/k_{//}]\pi^{1/2}[\xi_{2,d,1}+\xi_{2,d,0}^2\mathcal{Z}_{2,d}]$ in which we introduced the extra convention $\xi_{N,d,j}=[\omega-N\Omega +j k_{//}v_{d,//}]/[2^{1/2}v_{t,//}k_{//}]$.
Once the expressions for the dielectric response for a prescribed $\vec{k}$ are found, Bud\'{e}'s procedure requires to fit these functions. There are various ways of doing this. The brute-force method - exploiting $(R,Z,\varphi)$ as independent variables and recalling that the relevant parallel wave number involves both $\vec{k}$ and $\vec{k}'$ - is to perform a 4-dimensional fit in terms of $k_R$, $k_Z$, $k'_R$ and $k'_Z$ for a given set of toroidal mode numbers; fitting 1 direction at the time reveals itself to be more accurate than performing a single 4-D fit. 
   
\section{Upgrade to arbitrary distributions}

In \cite{DVE_arbitraryFo} or even more simply in \cite{AllFLRBudeDVE}, the procedure was illustrated on how - at the price of supplementary \textcolor{black}{computational} time - arbitrary distribution functions $F_o$ can be accounted for. Representing $F_o$ using a piece-by-piece linear-by-linear representation for an $F_o$ that is known on a sufficiently refined grid of points (either for an analytically known distribution or the numerical solution of a Fokker-Planck equation) the velocity space integral is broken up into a double sum of elementary integrals that can be solved by hand. The needed integrals are of the general form
\begin{equation}
\int_{v_{\perp,i}}^{v_{\perp,i+1}}dv_\perp \int_{v_{//,j}}^{v_{//,j+1}} dv_{//}\frac{[v_\perp-v_{\perp,i}]^m[v_{//}-v_{//,j}]^n}{k_{//}v_{//}+N\Omega-\omega}= 
\frac{1}{k_{//}} \mathrm{I}_{NR,m,\Delta_\perp} \mathrm{I}_{R,n,\Delta_{//}}        
\end{equation}
with $\mathrm{q}_{res}=[\omega-N\Omega-k_{//}v_{//,j}]/k_{//}$. The nonresonant integrals  $\mathrm{I}_{NR,m,\Delta}$ are
\begin{equation}\mathrm{I}_{NR,m,\Delta}=\frac{\Delta^{m+1}}{m+1}  
\end{equation}
while the resonant ones can be found from the recursion
\begin{equation} \mathrm{I}_{R,n,\Delta}=\mathrm{I}_{NR,n-1,\Delta}+\mathrm{q}_{res}\mathrm{I}_{R,n-1,\Delta} 
\end{equation}
with 
\begin{equation}\mathrm{I}_{R,0,\Delta}=\textcolor{black}{\ln} (\frac{\mathrm{q}_{res}-\Delta}{\mathrm{q}_{res}})=\textcolor{black}{\ln}(\frac{v_{//,j+1}-v_{//,res}}{v_{//,j}-v_{//,res}})=\textcolor{black}{\ln} \Big |\frac{v_{//,j+1}-v_{//,res}}{v_{//,j}-v_{//,res}} \Big |+i\pi [\tilde{\beta}_{j+1}-\tilde{\beta}_j]. 
\end{equation}
where $\tilde{\beta}_j$ is the argument of $v_{//,j}-v_{//,res}$. When the resonance is crossed and accounting for causality (replacing $\omega$ by $\omega+i\nu$ where $\nu$ is infinitesimal but positive as $\nu$ is physically associated with weak but finite collisionality) the logarithm picks up an imaginary part $i\pi |k_{//}|/k_{//}$: for positive $k_{//}$ the pole needs to be encircled from below and the arguments of $v_{//,j+1}-v_{//,res}$ and $v_{//,j}-v_{//,res}$ are $0$ and $-\pi$; for negative $k_{//}$ it has to be encircled from above so the arguments are $0$ and $+\pi$.

\section{Flux terms}

\textcolor{black}{It should be reminded that the here proposed variational method - suitable for finite element exploitation - does not express the wave equation in terms of the actual dielectric tensor but rather exploits an operator acting both on the electric field and the test function vector. Unless one wants to know what the wave fluxes are, the expression of the corresponding dielectric tensor (an operator solely acting on the electric field and not on the test function) thus strictly is not required. For a dielectric response operator truncated at low order finite Larmor radius corrections, the corresponding dielectric tensor expression can be readily derived by repeated partial integrations to remove all derivatives from the test function vector and transferring them to the electric field (see \cite{TOMCAT} for an example). In practice, the actual computation of this tensor quickly becomes cumbersome, in particular - as is the case for the Bud\'{e} method - if high order derivatives are needed. The total flux and its derivatives are, however, continuous so a suitable choice of equally smooth base functions ensures no net flux terms appear when assembling the linear system corresponding to the projection of the wave equation on these functions (again, see the above cited example for explicit expressions) eliminating the need for an explicit expression of the kinetic flux. By reverse engineering, the radial evolution of the total flux $\vec{S}$ can numerically be inferred from the power balance even when no explicit expression of the kinetic flux is available: By definition, $\nabla.\vec{S}+P_{abs}=0$ with $P_{abs}=\int d\vec{x} k_o^2 \vec{E}^*.\overline{\overline{\epsilon}}.\vec{E}/[i\omega \mu_o]$ (with $\overline{\overline{\epsilon}}$ the here adopted dielectric operator acting both on $\vec{E}$ and $\vec{F}$\textcolor{black}{, the latter being replaced by $\vec{E}$ when evaluating the absorption}), which can locally be evaluated to determine the total flux $\vec{S}$ crossing a magnetic surface, starting from the known source at the antenna and the wall while the integral over a magnetic surface infinitesimally close to the magnetic axis needs to approach zero. The radial component of the kinetic flux $\vec{S}_{kinetic}$ integrated over the magnetic surface can be found by subtracting the radial component of the magnetic surface integrated Poynting flux $\vec{S}_{Poynting}=\vec{E}^*\times \nabla \times \vec{E}/[i \omega \mu_o]$ from the total flux. A 1D illustration of this procedure is provided in \cite{TOMCAT-U1}.}

\section{Bud\'{e}'s method and parallel dynamics}
Since the angle $\Theta$ between the parallel and toroidal directions is small near the core where heating is typically taking place and because of the adopted choice of perpendicular unit vectors, adopting a truncated Taylor series expansion of the \textcolor{black}{plasma dispersion} function $\mathcal{Z}(\xi)$ is sometimes justified. Colestock and Kashuba used this approximation to illustrate how the poloidal field affects the wave dynamics \cite{Colestock}. The validity of such an expansion needs to be carefully checked, though. If allowed, this permits to make the poloidal k-component $k_\theta$ (and hence the $k_R$ and $k_Z$ components it depends on) explicit, the Taylor expansion together with the fits of the Kennel-Engelmann operator then in turn allowing to adopt the simplified fitting of the perpendicular dynamics part of the dielectric response functions $even$ in presence of a finite poloidal magnetic field: 
\begin{equation}
\mathcal{Z}(\xi) \approx \mathcal{Z}(\xi_{tor})+\frac{\partial \mathcal{Z}}{\partial k_R}k_R+\frac{\partial \mathcal{Z}}{\partial k_Z}k_Z+\frac{\partial^2 \mathcal{Z}}{\partial k_R^2}\frac{k_R^2}{2}+\frac{\partial^2 \mathcal{Z}}{\partial k_R\partial k_Z}k_Rk_Z+\frac{\partial^2 \mathcal{Z}}{\partial k_Z^2}\frac{k_Z^2}{2}.
\end{equation}
Here all partial derivatives are to be evaluated at $k_R=k_Z=0$ for a given cyclotron harmonic $N$
and $\xi=[\omega-N\Omega]/[2^{1/2}v_t k_{//}]$ while $\xi_{tor}=[\omega-N\Omega]/[2^{1/2}v_t k_{tor}].$ 
The partial derivatives in the above are determined by the chain rule, so e.g. 
\begin{equation}
\frac{\partial \mathcal{Z}(\xi)}{\partial k_\gamma} \approx \frac{d\mathcal{Z}}{d\xi}(\xi_{tor}) \frac{\partial \xi}{\partial k_{//}} \frac{\partial k_{//}}{\partial k_{\gamma}} 
\end{equation}
in which the first term is evaluated using $d\mathcal{Z}(\xi)/d\xi=-2[1+\xi\mathcal{Z} ]$, the second is simply $\partial \xi/\partial k_{//}=-\xi/k_{//}$ while the last is either $\partial k_{//}/ \partial k_R =-\textcolor{black}{\sin} \Theta \textcolor{black}{\sin} \alpha$ or $\partial k_{//}/ \partial k_Z =\textcolor{black}{\sin} \Theta \textcolor{black}{\cos}\alpha$;
$\alpha$ is the angle between $\vec{e}_\rho$ and $\vec{e}_R$. An obvious drawback of the above is that the parallel dynamics is not properly modeled: the $k_{//}$- up- or downshift due to the presence of the poloidal field is inadequately modeled since the $k_{\theta}$-dependent terms have been removed from the argument of the plasma dispersion function, an effect that is particularly visible when the toroidal mode number is small and/or the poloidal component of the wave vector significant so that the poloidal corrections to $k_{//}$ are important. 

\textcolor{black}{The fact that the parallel gradient remains a \textit{differential} operator - as opposed to an algebraic one - is a drawback of the Bud\'{e} procedure since the above computed simple Taylor series expansion is thus not necessarily a good representation.} Bud\'{e} himself illustrated his elegant and powerful procedure in absence of the poloidal field and in 1D \cite{Bude}. He stated that exploitation of his procedure beyond 1D would actually yield \textcolor{black}{computational} benefits rather than extra bottlenecks compared to solving the full integro-differential equation. In absence of poloidal field effects it is certainly true that his procedure is indeed much less \textcolor{black}{time-consuming}. In presence of a finite poloidal field this is no longer necessarily the case. In case a procedure could be found that decouples the parallel from the perpendicular dynamics (i.e. a procedure that directly reasons in terms of $x_{//}$ to capture the isolated parallel dynamics in some way) a further speed-up could possibly be realised. 
\textcolor{black}{D}eriving the hot plasma conductivity tensor for a tokamak, Svidzinski \cite{Svidzinski} explored a route to include the parallel dynamics by expressing the electric field at the position $(R',Z')$ in the orbit integral yielding the dielectric response in terms of its values at the grid points $(R_i,Z_j)$ later adopted for the actual numerical solving of the wave equation. Adopting a sufficiently refined grid, this allows him to locally represent the $\vec{E}(R',Z')$ using a low order polynomial (he uses second order Lagrange polynomials) and define a nonlocal conductivity tensor accounting for the poloidal field. He performs the needed integrals semi-analytically and labels the resulting procedure as computationally expensive, requiring parallel processing.  

Applying Svidzinski's philosophy to Bud\'{e}'s method as a means to seek to push down the \textcolor{black}{computational} requirements by performing some of the integrals by hand amounts to introducing a grid-defined local approximation of the integrand: Whereas Svidzinski's approach makes a Taylor series expansion of the electric field and expresses the local contribution in terms of the electric field values at the 4 corner points of a local finite element, the here adopted approach cannot make such a distinction since it dominantly operates in $\vec{k}$ space so the electric field (here actually its Fourier component $\vec{E}_{\vec{k}}$) is common to all 4 corner points; it is via the \textcolor{black}{exponential factor $exp[i (\vec{k}-\vec{k}').\vec{x}]$ } that the location where $\vec{E}$ is evaluated is specified. Adopting a sufficiently refined $(R,Z)$ grid and realising all functions are locally smooth (the least smooth function appearing being the causality smoothed logarithm when performing the parallel velocity integral in the case of an arbitrary distribution function and requiring the most tight gridding) we can locally perform a Taylor series expansion in $\vec{x}$ for \textit{fixed} \textcolor{black}{$(k_R,k_Z)$}, however: the reason why a Taylor series expansion in the discussion in this section so far 
is not always suitable is because of the wide range of $\vec{k}$, not the mild variation due to local changes of the equilibrium quantities or the rotation matrix $\overline{\overline{R}}$. For any given $\vec{k}$ a proper grid can be chosen to ensure the variation of the integrand is mild in the local finite element. \textcolor{black}{If the shortest wavelength wave expected in the solution has a wave vector with magnitude $k_{max}=|\vec{k}|$ then the grid spacings $\Delta R = R_{i+1}-R_i $ and $\Delta Z = Z_{j+1}-Z_j $ should be chosen such that both $k_{max} \Delta R << 1 $ and $k_{max} \Delta Z <<1 $. These conditions whatsoever need to be satisfied to ensure the finite element procedure is sufficiently accurate for the expected wavelengths. } The type of integrals to perform for a given toroidal mode number is 
\begin{equation}
\int_{R_i}^{R_{i+1}}  \int_{Z_j}^{Z_{j+1}} dRdZ \int dk_Rdk_Z \int dk'_Rdk'_Z e^{[i(\vec{k}-\vec{k}\textcolor{black}{'}).\vec{x}]} F_{\vec{k}',\alpha}^*\mathcal{P}(\zeta,\zeta',\psi,\psi',R,Z)\mathcal{Q}(k_{//},R,Z) 
E_{\vec{k},\beta} 
\end{equation}
in which the first 2 integrals are on a small finite element, $\zeta$ is the argument of the Bessel functions ($\zeta=k_\perp\rho_L$) in which $k_\perp(k_R,k_Z,R,Z)$ and similar for $k_{//}$, $k'_\perp$ and $k'_{//}$ but not for $\psi$ nor $\psi'$. Inside the finite element we can adopt the local variables $\zeta_R$ and $\zeta_Z$ varying from 
\textcolor{black}{ $-0.5$ to $0.5$\textcolor{black}{.}} 
For a sufficiently refined grid, we can then represent the various functions in each interval making use of a suitable set of low order polynomial (finite-element) base functions
e.g. 
\begin{equation} \textcolor{black}{\mathcal{P}}
\approx \mathcal{P}_{i,j}+\frac{\partial \mathcal{P} }{\partial R}\Big |_{i,j} \textcolor{black}{\Delta R} \zeta_R+\frac{\partial \mathcal{P} }{ \partial Z}\Big |_{i,j} \textcolor{black}{\Delta Z}\zeta_Z+\frac{\partial^2 \mathcal{P} }{\partial R \partial Z}\Big |_{i,j} \textcolor{black}{\Delta R} \textcolor{black}{\Delta Z}\zeta_R\zeta_Z+...  
\end{equation}
and similar for $\mathcal{Q}$ as well as the exponential factor, the partial derivatives in which make extra powers of $k_R$ and $k_Z$ appear via partial derivatives of $\zeta=k_\perp \rho_L$ (Larmor radius corrections) and $k_{//}$ (parallel gradient corrections) on $R$ and $Z$ for given $k_R$ and $k_Z$\textcolor{black}{.} 
\textcolor{black}{In particular we get expressions involving the plasma dispersion function
$$\mathcal{Z} (\xi) \approx \mathcal{Z}(\xi_{i,j})+\frac{\partial \mathcal{Z}}{\partial \xi} \Big [  \frac{\partial \xi}{\partial R} \Delta R \zeta_R + \frac{\partial \xi}{\partial Z} \Delta Z \zeta_Z \Big ] $$
in which the second and third term in the right hand side is - by construction - an as small correction as required.} 
\textcolor{black}{The} final result \textcolor{black}{after integrating over $\zeta_R$, $\zeta_Z$ (which just requires integrating low order polynomials over the small intervals) and performing the inverse Fourier transform - after adopting Bud\'{e}'s procedure to fit the various functions to a high order polynomial -} is an expression solely in terms of the electric field, the test function vector and their derivatives at a series of reference points \textcolor{black}{$(R_{mid},Z_{mid})$}.

The advantage of considering a volume rather than an individual point when evaluating the dielectric response in $\vec{k}$ space is that this has a smoothing effect and hence potentially allows reducing the order of the polynomials needed to fit the response. This can most easily be illustrated when considering the integration of the parallel integral for arbitrary $F_o$ (see section 5), which also represents the toughest case where the resonance crossing - represented by the argument of the logarithm crossing zero - locally yields a discontinuous jump of the imaginary part of the logarithm, traditionally smoothed by introducing a finite collisionality. The relevant type of integrals is of the form
\begin{equation}
\int d\zeta_R d\zeta_Z \zeta_R^m \zeta_Z^n \textcolor{black}{\ln}(\frac{[ v_{//,i+1}-v_{//,res} ]}{[ v_{//,i}-v_{//,res}]})  
\end{equation}   
which can be evaluated repeatedly using the elementary expression \cite{Abramowitz}
\begin{equation}
\int dz z^m \textcolor{black}{\ln} z=\frac{z^{m+1}}{m+1}[\textcolor{black}{\ln} z-\frac{1}{(m+1)}].  
\end{equation}

\textcolor{black}{Mathematically, Bud\'{e}'s fitting-based method is on firm ground but it requires high-order derivatives, the accuracy of the numerical evaluation of which may be difficult. 
}
Note that because of this fitting \textcolor{black}{- or (provided all degrees of freedom are exploited) equivalent differencing - } procedure the road to 3D application is - mathematically speaking - immediately open, be it that extensive \textcolor{black}{computational} time needs to be devoted to compute the needed functions for a series of $\vec{k}$ values at all grid positions, allowing to find the required fits. As already mentioned, the bottlenecks deciding on the usefulness of the Bud\'{e} procedure are expected to be of practical (\textcolor{black}{computer} time and memory requirements) rather than of mathematical nature. Rather than a single toroidal mode number (or a set of coupled toroidal mode numbers and retaining the coupling between them, as was mentioned earlier), a polynomial fit now also involving the toroidal mode number then appears and can - just like for the other $\vec{k}$'s - be transformed back to higher order derivatives in the toroidal direction. At that stage, not even the differential operators accounting for the curvature are needed: exploiting the rotation matrices, all can simply be expressed in terms of the basic cartesian $(X,Y,Z)$ coordinates rather than the here adopted $(R(X,Y),Z,\varphi(X,Y))$ to account for the toroidal curvature and be written exploiting the basic Cartesian differential operators. \textcolor{black}{For example, applying the Bud\'{e} equivalent of Svidzinski's procedure now involves a set of given fixed $(k_X,k_Y,k_Z)$ rather than the fixed cylindrical $(k_R,k_Z)$ used in the present \textcolor{black}{paper.} }

\section{The wave equation: practical form}

In variational form and assuming the base functions exploited by the finite element exploitation are properly chosen so that internal flux terms are unnecessary, the wave equation can be reduced to
\begin{equation}
  2\pi \int dR dZ R \Bigg [  [k_o^2 \vec{F}^*. \overline{\overline{\epsilon}}.\vec{E} \Big ]  -( \nabla \times \vec{F})^*.(\nabla \times \vec{E}) \Bigg ]=0 
\label{eq : BasicFinalEq} 
\end{equation} 
where 
\begin{equation}
\vec{F}^*.\overline{\overline{\epsilon}}.\vec{E}=\vec{F}^*.\overline{\overline{\mathcal{P}}}_0.\vec{E}-\sum_{I,J,K,L}  (-i)^{I+J-K-L} \frac{\partial^{K+L} \vec{F}^* }{\partial R^K \partial Z^L}.\overline{\overline{\mathcal{P}}}_{1,I,J,K,L}.\frac{\partial^{I+J}\vec{E}}{\partial R^I \partial Z^J}. 
\label{eq : BasicFinalEqCoef} 
\end{equation}
which can be expressed in terms of $(E_R,E_Z,E_\varphi)$ when including the rotation matrix $\overline{\overline{\mathcal{R}}}$ and its inverse or in terms of $(E_{\perp,1},E_{\perp,2},E_{//})$  when omitting it.

In matrix form, the expression for $( \nabla \times \vec{F})^*.(\nabla \times \vec{E})$ in terms of the components parallel and perpendicular to $\vec{B}_o$ is
\begin{equation}
( \nabla \times \vec{F})^*.(\nabla \times \vec{E})=  
\Bigg (\begin{array}{ccc} F_{\perp,1}^* & F_{\perp,2}^* & F_{//}^*  \end{array}\Bigg ) . \Bigg [ \overline{\overline{C}}_{oo} + \overleftarrow{\partial}_R \overline{\overline{C}}_{Ro}  + \overleftarrow{\partial}_Z \overline{\overline{C}}_{Zo}+ \overline{\overline{C}}_{oR} \overrightarrow{\partial}_R + \overline{\overline{C}}_{oZ} \overrightarrow{\partial}_Z  $$
$$ + \overleftarrow{\partial}_R \overline{\overline{C}}_{RR} \overrightarrow{\partial}_R  + \overleftarrow{\partial}_R \overline{\overline{C}}_{RZ} \overrightarrow{\partial}_Z + \overleftarrow{\partial}_Z \overline{\overline{C}}_{ZR} \overrightarrow{\partial}_R + \overleftarrow{\partial}_Z \overline{\overline{C}}_{ZZ} \overrightarrow{\partial}_Z \Bigg  ] .  \Bigg ( \begin{array}{c} E_{\perp,1} \\ E_{\perp,2} \\ E_{//}  \end{array} \Bigg )   
\end{equation}
for which the explicit expression of the various matrices can be found in \textcolor{black}{\cite{VanEester_2DBude_LabReport}}.
The arrows above the partial differential operators in the above indicate in which direction the operator acts, to the left on the test function vector or to the right on the electric field.
Like the plasma term we write the volume terms as a sum of contributions referring to the various derivatives
\begin{equation}
( \nabla \times \vec{F})^*.(\nabla \times \vec{E})=\sum_{I,J,K,L}  \frac{\partial^{K+L} \vec{F}^* }{\partial R^K \partial Z^L}.\overline{\overline{\mathcal{C}}}_{I,J,K,L}.\frac{\partial^{I+J}\vec{E}}{\partial R^I \partial Z^J}  
\end{equation}
where all introduced notations can be identified by looking at the corresponding matrix expressions provided. The whole wave equation can now be written in matrix form as
\begin{equation}
2\pi \int dR dZ R \Bigg [  \sum_{I,J,K,L}  (-i)^{I+J-K-L} \frac{\partial^{K+L} \vec{F}^* }{\partial R^K \partial Z^L}. \Bigg  [ k_o^2  [\overline{\overline{\mathcal{P}}}_{0,I,J,K,L}-\overline{\overline{\mathcal{P}}}_{1,I,J,K,L}]   -\overline{\overline{\mathcal{C}}}_{I,J,K,L} \Bigg ] . \frac{\partial^{I+J}\vec{E}}{\partial R^I \partial Z^J} \Bigg ] =0. 
\end{equation}

When exploiting nonlinear regression to define a proper fit in $\vec{k}$-space allowing to recast the integro-differential wave equation into a higher order partial differential equation following the technique proposed by Bud\'{e}, the choice of the order of the polynomial is a balance between ensuring a suitably correct fit (which suggests choosing $M$ as large as possible) and avoiding the reigning partial differential equation has unpractically high partial derivatives (which suggests choosing $M$ as modest as possible). \textcolor{black}{Nonlinear regression minimises the summed "distance" between a set of prescribed values of a function  $f(\zeta)$ to a polynomial approximation $f_{fit}=\sum_{m=0}^M f_m (\zeta-\zeta_{ref})^m$. This philosophy can easily be extended to multiple dimensions, finding the fits for 1 dimension at the time. Practical exploitation will require optimisation to push down the required computational effort while guaranteeing sufficient accuracy. 
} 

\section{Reduced finite Larmor radius expansion for Maxwellian plasmas}

The expressions provided in this paper allow to treat the full integro-differential wave equation. To cut down the amount of algebra in the preparatory phase preceding the actual solving of the wave equation 
but still being able to treat the key physics, the expressions of the dielectric response have often been simplified by exploiting truncated Taylor series expansions of the various Bessel functions at terms of second order in the Larmor radius. Thinking about applying Bud\'{e}'s method beyond the so far existing 1D explorations, one may wonder whether it makes sense to include an intermediate step and adopt truncated finite Larmor radius expressions rather than accounting for the fully kinetic description valid at any temperature and for any wavelength. Whereas traditionally (see e.g. \cite{Brambilla}) the dielectric tensor itself was expanded in terms of $k_\perp\rho_L$ (with $k_\perp$ is the perpendicular wave vector component and $\rho_L$ is the Larmor radius), the variational approach adopted here leans on the Kennel-Engelmann operator which appears twice in the expression of the dielectric response, once acting on the electric field $\vec{E}$ and once acting on the test function vector $\vec{F}$. In \cite{TOMCAT}, the 1D version of the wave equation in absence of poloidal field effects was presented leaning on this approach. Since it retained up to second order finite Larmor radius terms of the Taylor series expansion of the Kennel-Engelmann operator, it yielded a 12th order differential equation rather than the traditional 6th order system. In the present paper 2D application is prepared and the poloidal field effects are no longer neglected. As a consequence, the (2D) expressions presented in the Appendix of \cite{TOMCAT} can not directly be exploited but the corresponding Taylor series expansion in the perpendicular directions still holds. Conform with the here adopted notation (in which the sign of the cyclotron harmonic index $N$ has been flipped w.r.t. the expressions in the original paper) the dielectric response in terms of the $(+,-,//)$ components of the wave equation and the electric field read

$$\overline{\overline{P}}_{--} \approx [-_+k^{'2}] ^*\frac{3}{\textcolor{black}{8}}\tilde{\tilde{A}}_1 [-k_+]^2  + [-i_+k']^*\tilde{A}_0 [-ik_+] + A_{-1} + [\textcolor{black}{-_\perp k^{'2}}]^* \tilde{A}_{-1} + \tilde{A}_{-1}  [ \textcolor{black}{-k^2_\perp}]  $$
$$+ [\textcolor{black}{-_\perp k^{'2}}]^*\frac{3}{2}\tilde{\tilde{A}}_{-1} [\textcolor{black}{-k^2_\perp}] + [-i_-k']^*\tilde{A}_{-2} [-ik_-] + [-_-k^{'2}]^*\frac{3}{\textcolor{black}{8}}\tilde{\tilde{A}}_{-3} [-k_-^2]  $$ 
$$\overline{\overline{P}}_{-+} \approx -\Bigg [  [-_+k^{'2}]^* \textcolor{black}{\frac{1}{2}} \tilde{A}_1 +[-_+k^{'2}]^*\frac{3}{\textcolor{black}{4}}\tilde{\tilde{A}}_1 [\textcolor{black}{-k^2_\perp}] + [-i_+k']^* \tilde{A}_0 [-ik_-] + \textcolor{black}{\frac{1}{2}}\tilde{A}_{-1} [-k_-^2] + [\textcolor{black}{-_\perp k^{'2}}]^*\frac{3}{\textcolor{black}{4}}\tilde{\tilde{A}}_{-1} [-k_-^2] \Bigg ] $$ 
$$\overline{\overline{P}}_{-//}\approx i \Bigg [ [-_+k^{'2}]^* \textcolor{black}{\frac{1}{2}} \tilde{\tilde{B}}_1[-ik_+] + [-i_+k']^* \tilde{B}_0 + [-i_+k']^*\tilde{\tilde{B_0}} [\textcolor{black}{-k^2_\perp}]+\tilde{B}_{-1}[-ik_-] +[ \textcolor{black}{-_\perp k^{'2}} ]^* \tilde{\tilde{B}}_{-1} [-ik_-] $$
$$+[-i_-k']^* \textcolor{black}{\frac{1}{2}}\tilde{\tilde{B}}_{-2} [-k_-^2]\Bigg ]  $$
$$\overline{\overline{P}}_{+-} \approx -\Bigg [  \textcolor{black}{\frac{1}{2}} \tilde{A}_1[-k_+^2] +[\textcolor{black}{-_\perp k^{'2}}]^*\frac{3}{\textcolor{black}{4}}\tilde{\tilde{A}}_1 [-k_+^2] + [-i_-k']^* \tilde{A}_0 [-ik_+] + [-_-k^{'2}] \textcolor{black}{\frac{1}{2}} \tilde{A}_{-1}  + [_-k ^{'2}]^* \frac{3}{\textcolor{black}{4}} \tilde{\tilde{A}}_{-1} [\textcolor{black}{-k^2_\perp}] \Bigg ] $$ 
$$\overline{\overline{P}}_{++} \approx  [-_+k^{'2}]^*\frac{3}{\textcolor{black}{8}}\tilde{\tilde{A}}_{3} [-k_+^2] + [-i_+k']^*\tilde{A}_{2} [-ik_+] + A_{1} + [\textcolor{black}{-_\perp k^{'2}}]^* \tilde{A}_{1} + \tilde{A}_{1}  [ \textcolor{black}{-k^2_\perp}]  $$
$$+ [\textcolor{black}{-_\perp k^{'2}}]^*\frac{3}{2}\tilde{\tilde{A}}_{1} [\textcolor{black}{-k^2_\perp}] + [-i_-k']^*\tilde{A}_0 [-ik_-]  +  [-_-k^{'2}]^*\frac{3}{\textcolor{black}{8}}\tilde{\tilde{A}}_{-1} [-k_-]^2   $$ 
$$\overline{\overline{P}}_{+//}\approx - i \Bigg [  +[-i_+k']^* \textcolor{black}{\frac{1}{2}}\tilde{\tilde{B}}_{2} [-k_+^2]\Bigg ] +\tilde{B}_{1}[-ik_+] +[ \textcolor{black}{-_\perp k^{'2}} ]^* \tilde{\tilde{B}}_{1} [-ik_+] + [-i_-k']^* \tilde{B}_0 + [-i_-k']^*\tilde{\tilde{B_0}} [\textcolor{black}{-k^2_\perp}] $$
$$+ [-_-k^{'2}]^*\textcolor{black}{\frac{1}{2}}\tilde{\tilde{B}}_{-1}[-ik_-]  $$
$$\overline{\overline{P}}_{//-}\approx -i \Bigg [ [-i_+k']^* \textcolor{black}{\frac{1}{2}} \tilde{\tilde{B}}_1[-k_+^2] +\tilde{B}_0 [-ik_+]  + [\textcolor{black}{-_\perp k^{'2}}]^*\tilde{\tilde{B_0}} [-ik_+]+[-i_-k']^* \tilde{B}_{-1}+[ -i_-k']^* \tilde{\tilde{B}}_{-1} [\textcolor{black}{-k^2_\perp}] $$
$$+[-_-k^{'2}]^* \textcolor{black}{\frac{1}{2}} \tilde{\tilde{B}}_{-2} [-ik_-]\Bigg ]  $$
$$\overline{\overline{P}}_{//+}\approx i \Bigg [ +[-_+k^{'2}]^* \textcolor{black}{\frac{1}{2}}  \tilde{\tilde{B}}_{2} [-ik_+] +[-i_+k']^*\tilde{B}_{1} +[ -i_+k' ]^* \tilde{\tilde{B}}_{1} [\textcolor{black}{-k^2_\perp}] + [\textcolor{black}{-_\perp k^{'2}}]^*\tilde{\tilde{B_0}} [-ik_-] +  \tilde{B}_0[-ik_-]  $$
$$+ [-i_-k']^* \textcolor{black}{\frac{1}{2}} \tilde{\tilde{B}}_{-1}[-k_-^2] \Bigg ]  $$
$$\overline{\overline{P}}_{// //}\approx [-_+k^{'2}]^* \textcolor{black}{\frac{1}{4}} \tilde{\tilde{C}}_2[-k_+^2] + [-i_+k']^* \tilde{C}_1 [-ik_+] + \textcolor{black}{2} C_0 +[\textcolor{black}{-_\perp k^{'2}}]^* \tilde{C}_0 +\tilde{C}_0 [\textcolor{black}{-k^2_\perp}] $$
\begin{equation}
+   [\textcolor{black}{-_\perp k^{'2}}]^* \tilde{\tilde{C}}_0 [\textcolor{black}{-k^2_\perp}] + [-i_-k']^* \tilde{C}_{-1} [-ik_-] + [-_-k^{'2}]^* \textcolor{black}{\frac{1}{4}} \tilde{\tilde{C}}_{-2}[-k_-^2] 
\end{equation}

in $\vec{k}$-space and where $_{...}k'$ refers to wave vector components of the test function and $k_{...}$ to those of the RF electric field, and similar for $_\perp k^{'2}$ and $k^2_\perp$; the plasma contribution to the wave equation is then $\vec{F}^*.\overline{\overline{P}}.\vec{E}$. 
This can immediately be recast in terms of the $(\perp_1,\perp_2,//)$ components 
via the transformation 

\begin{equation}
\Bigg ( \begin{array}{c} w_+ \\ w_- \\ w_{//}\end{array}\Bigg )= \Bigg ( \begin{array}{ccc} 1 & i & 0 \\ 1 & -i & 0 \\ 0 & 0 & 1 \end{array}\Bigg ). \Bigg ( \begin{array}{c} w_{\perp,1} \\ w_{\perp,2} \\ w_{//}\end{array}\Bigg )  
\end{equation}

and then further via the already discussed $\overline{\overline{\mathcal{R}}}$ to the cylindrical $(R,Z,\varphi)$. The coefficients in the above are

$$A_N=k_0^2 \frac{\omega_p^2 \mathcal{Z}_N}{2^{3/2}v_tk_{//}\omega}$$
$$B_N=k_0^2 \frac{\omega_p^2 (1+\xi_N \mathcal{Z}_N)}{2v_tk_{//}\omega}$$
$$C_N=k_0^2 \frac{\omega_p^2 \xi_N(1+\xi_N \mathcal{Z}_N)}{2^{\textcolor{black}{1}/2}v_tk_{//}\omega}$$
$$\tilde{A}_N=\rho_L^2A_N$$
$$\tilde{B}_N=\rho_LB_N$$
$$\tilde{C}_N=\rho_L^2C_N$$
$$\tilde{\tilde{A}}_N=\rho_L^4A_N$$
$$\tilde{\tilde{B}}_N=\rho_L^3B_N$$
\begin{equation}
\tilde{\tilde{C}}_N=\rho_L^4C_N  
\end{equation}

in which $\mathcal{Z}_N$ is the plasma dispersion function with argument $\xi_N$. In case the impact of the poloidal field is omitted, $k_{\perp,1}=k_R$ and $k_{\perp,2}=k_Z$ so it suffices to replace the above $k_+$ by $-i \nabla_+$ and $k_-$ by $-i \nabla_-$ acting to the right on $\vec{E}$ and similarly to substitute $_+k'$ by $-i _+\nabla$ and $_-k'$ by $-i _-\nabla$ acting to the left on $\vec{F}$ to obtain the expressions of the 2D version of the dielectric response operator adopted in \cite{TOMCAT}. But when the poloidal field is accounted for there is also $k_R$ and $k_Z$ dependence in $k_{//}$ so this simple procedure no longer holds.

Avoiding the complication of retaining the full expression of the Kennel-Engelmann operators in the dielectric tensor by adopting the here presented truncated Taylor series expansion constitutes an intermediate way to check if the Bud\'{e} method - or its upgrade including the Bud\'{e} variant of the Svidzinski approach - has potential in the here presented form accounting for the finite poloidal magnetic field, which was neglected in \cite{TOMCAT}. If this test proves unsuccessful, one possible alternative is to return to the $(\rho,\theta,\varphi)$ representation, which makes the parallel gradient an algebraic operator rather than a differential one ($\partial ... /\partial \theta=im ...$ where m is the poloidal mode number of the electric field or the test function vector) but forces one to cope with the mathematical singularity at the magnetic axis. 

\section{Future plans and conclusions}

In the present paper, the semi-analytical expressions required to solve the 2D all-FLR integro-differential wave equation reigning the wave dynamics in the ICRH domain while adopting the Bud\'{e} method have been derived. Including all finite Larmor radius effects normally requires solving an integro-differential equation. Whereas the usual procedure is to rely on a Taylor series expansion of the Bessel functions appearing in the expression of the dielectric response in $\vec{k}$-space - a procedure strictly speaking limiting the application to wave modes that do not violate the smallness assumption of $k_\perp\rho_L$ (with $k_\perp$ the perpendicular wave vector component and $\rho_L$ the Larmor radius) - Bud\'{e} proposed to solve that equation as a high order partial differential equation by invoking a \textit{fitting} procedure allowing to catch the dependence in $\vec{k}$ space for both long \textit{and} short wavelength modes but still keeping the order of the fitting polynomial (and hence the order of the differential operators when transforming back to $\vec{x}$ space) modest. 

Future work involves implementing the presented procedure to actually solve the 2D wave equation. Although the proposed partial differential equation is mathematically well defined, a critical assessment will be needed to make sure the fitting and solving procedure is sufficiently fast and accurate to make exploitation practical. 
\textcolor{black}{The} fact that the poloidal magnetic field needs to be accounted for makes that not only the dynamics perpendicular but also that parallel to the static magnetic field needs to be accounted for in the fitting procedure\textcolor{black}{, making the parallel gradient an actual differential operator}. This may require polynomial fits of too high order to be practical, in particular when the argument of the \textcolor{black}{plasma dispersion} functions passes zero or infinity\textcolor{black}{.}

\textcolor{black}{Provided} the polynomial order of the fits in $\vec{k}$ space can be kept sufficiently modest to stay practical, the Bud\'{e} method allows to use off-the-shelf finite element solvers such as \textcolor{black}{MFEM} \cite{MFEM}, exploiting grid refinement techniques and exploring the benefits of higher order polynomial representations of the base functions exploited in finite element equation solving schemes.

\end{document}